\documentclass[aps,pra,twocolumn,groupedaddress,showpacs]{revtex4}

\usepackage{subfigure,epsfig,amsfonts,amsmath,graphicx,dcolumn,bm,array,caption}

\begin{document}

\title{Analysis of Photonic Quantum Nodes Based on Single Photon Raman Interaction}

\author{Serge Rosenblum$^1$}
\author{Adrien Borne$^1$}
\author{Barak Dayan$^1$}
\affiliation{$^1$Department of Chemical Physics, Weizmann Institute of Science, Rehovot 76100, Israel}

\date{\today}

\begin{abstract}
The long-standing goal of deterministically controlling a single photon using another was recently realized in various experimental settings. Among these, a particularly attractive demonstration relied on single-photon Raman interaction (SPRINT) in a three-level $\Lambda$-system coupled to a single-mode waveguide.
Beyond the ability to control the direction of propagation of one photon by the direction of another photon, this scheme has the potential to perform as a passive quantum memory and a universal quantum gate.
Relying on interference, this all-optical, coherent scheme requires no additional control fields, and can therefore form the basis for scalable quantum networks composed of passive quantum nodes that interact with each other only with single photon pulses.
Here we present an analytical and numerical 
study of SPRINT, and
characterise its limitations and the parameters for optimal operation. Specifically, we study the effect of losses and the presence of multiple excited states. In both cases we discuss strategies for restoring the high fidelity of the device's operation.

\end{abstract}
\pacs{42.50.Ct, 42.50.Pq, 42.50.Ex, 42.50.Dv}
\maketitle
	
\section{\label{sec:level1}Introduction}

Optical photons are widely considered a prime candidate for the transmission of quantum information from one material quantum node, responsible for processing and storage, to another~\cite{kimble2008}.
A prerequisite for such a hybrid quantum system is a reliable interface between its material and optical components.
Such an interface also has the prospect of creating an effective interaction between different photons, and may therefore act as a platform for all-optical quantum processing.
Recently, much experimental progress has been made towards this goal.
A switch in which a single control photon strongly modifies the attenuation of signal photons has been achieved by electromagnetically induced transparency and Rydberg blockade~\cite{chen2013,baur2014}.
The powerful scheme proposed by Duan and Kimble~\cite{duan2004}, in which microwave or Raman beams are used to create a single-atom interferometer that responds to the presence of one photon in the cavity mode, has been used to experimentally realize nondestructive detection of optical photons~\cite{reiserer2013}, a phase gate~\cite{tiecke2014}, and a quantum gate between photons and a single atom~\cite{reiserer2014}.

In this work we focus on the recent demonstrations of a single photon router~\cite{shomroni2014} and a single photon extractor~\cite{rosenblum2015}, which were based on single-photon Raman interaction (SPRINT) with a three-level $\Lambda$-atom coupled to a single-mode waveguide.
Originally proposed by Pinotsi and Imamoglu as a deterministic absorber of a single photon~\cite{pinotsi2008} and a quantum memory~\cite{pinotsi2008,lin2009}, the mechanism of SPRINT was shown by Koshino et al. to be able to implement a $\sqrt{\textrm{SWAP}}$ universal quantum gate~\cite{koshino2010,gea2012} as well. Later studies focused on the harnessing of this scheme for photon-photon interactions, such as photon routing and single-photon extraction~\cite{rosenblum2011}, and single-photon addition~\cite{gea2013}.
\begin{figure}[b!]
\begin{center}
\includegraphics[width=0.4\textwidth]{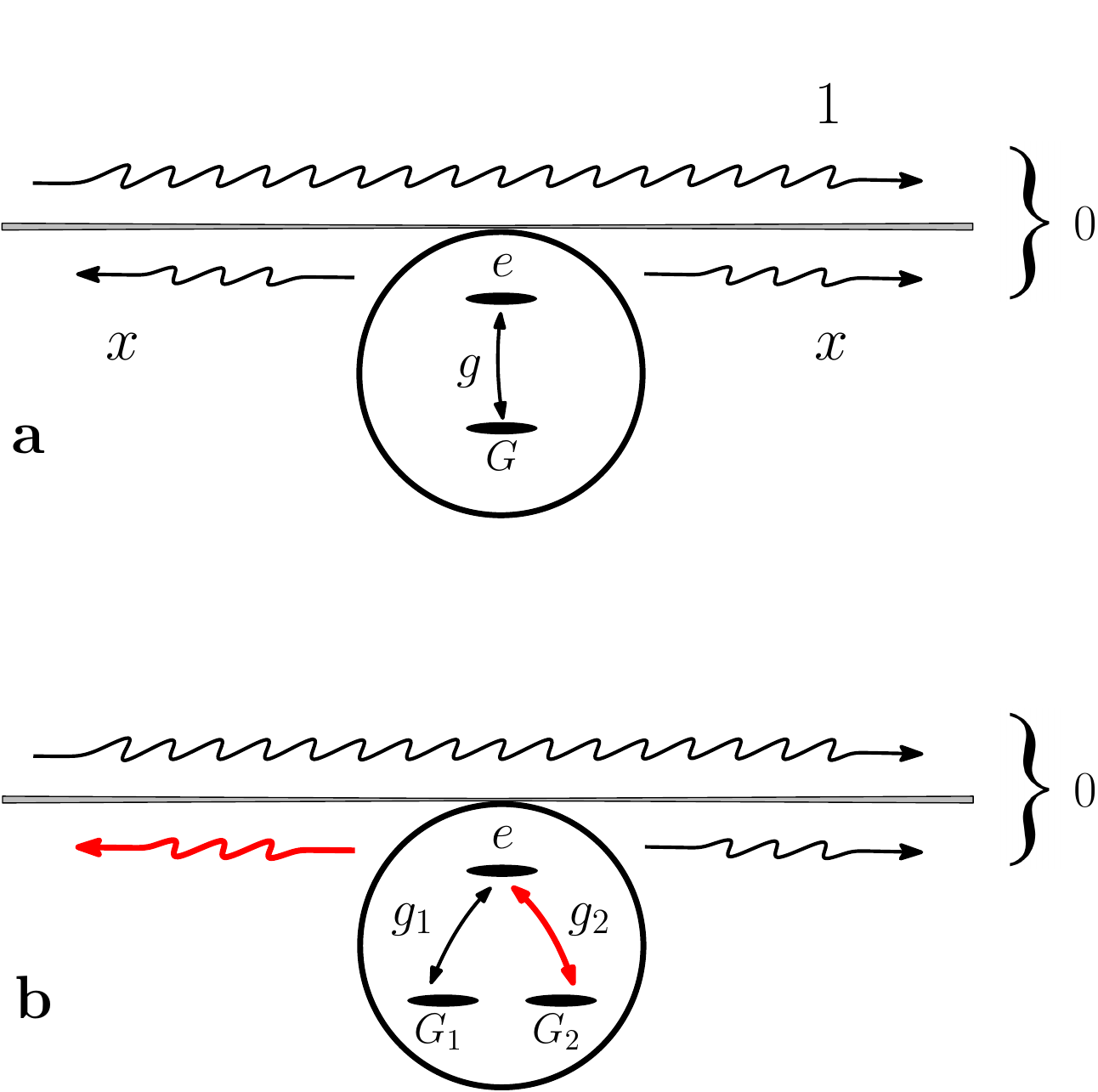}
\end{center}
\caption{(a) A two-level system coupled to a single-mode waveguide resulting in complete reflection of the probe. (b) A three-level $\Lambda$-system in which reflection of a single photon enforces a Raman transfer of the atom.}
\label{fig:principles_new}
\end{figure}

The underlying mechanism of SPRINT is evident if we first consider the simpler case of a lossless two-level system coupled to a single-mode waveguide (Fig. \ref{fig:principles_new}a).
When a weak resonant probe is sent through the waveguide at an amplitude of $1$, and the atom reacts by radiating into both directions with an amplitude $x$, conservation of energy dictates that $|x|^2+|1+x|^2=1$, and hence the only nontrivial solution is $x=-1$. As a result, the atom must reflect all incoming light, due to destructive interference between the probe and the forward radiation of the atom. This is in fact the very mechanism responsible for the complete reflection from a metallic mirror: the induced displacement of the free charges lags by a $\pi/2$ phase compared to the driving probe, and the field radiated by the charges lags by another $\pi/2$, leading to complete destructive interference with the probe in the forward direction, and hence to its reflection~\cite{pi_note}.
In the three-level version of this scenario, each `leg' of the $\Lambda$-system is coupled to a different direction of the waveguide (as depicted in Fig.~\ref{fig:principles_new}b). This modification does not change the fact that if the amplitude of radiation of the atom to both directions is equal, the result is total destructive interference in the forward direction. Yet, in contrast to the two-level case, reflection of the incoming photon results in Raman transfer of the atom to the second ground state. This three-level system therefore behaves like a single-photon mirror with a memory.
Note that SPRINT also occurs in a single-sided Fabry-Pérot resonator, in which each `leg' of the $\Lambda$-system is coupled to a mode of different frequency or polarization.\\

In the following section we describe the implementation of SPRINT in Ref.~\cite{shomroni2014}, and in Section~\ref{sec:level2} we provide the theoretical framework for modeling its operation. In Section~\ref{sec:level3} we describe the consequences of multiple excited states through which the passage can take place, and propose a way for restoring its performance. Finally, in Section~\ref{sec:level4} we describe the results of a full simulation taking into account various effects present in the practical implementation of SPRINT in Ref.~\cite{shomroni2014}, and discuss the optimal choice of parameters.\\

\section{\label{sec:level1}Realization of SPRINT with a single atom coupled to a WGM resonator}

\begin{figure}
\begin{center}
\includegraphics[width=0.4\textwidth]{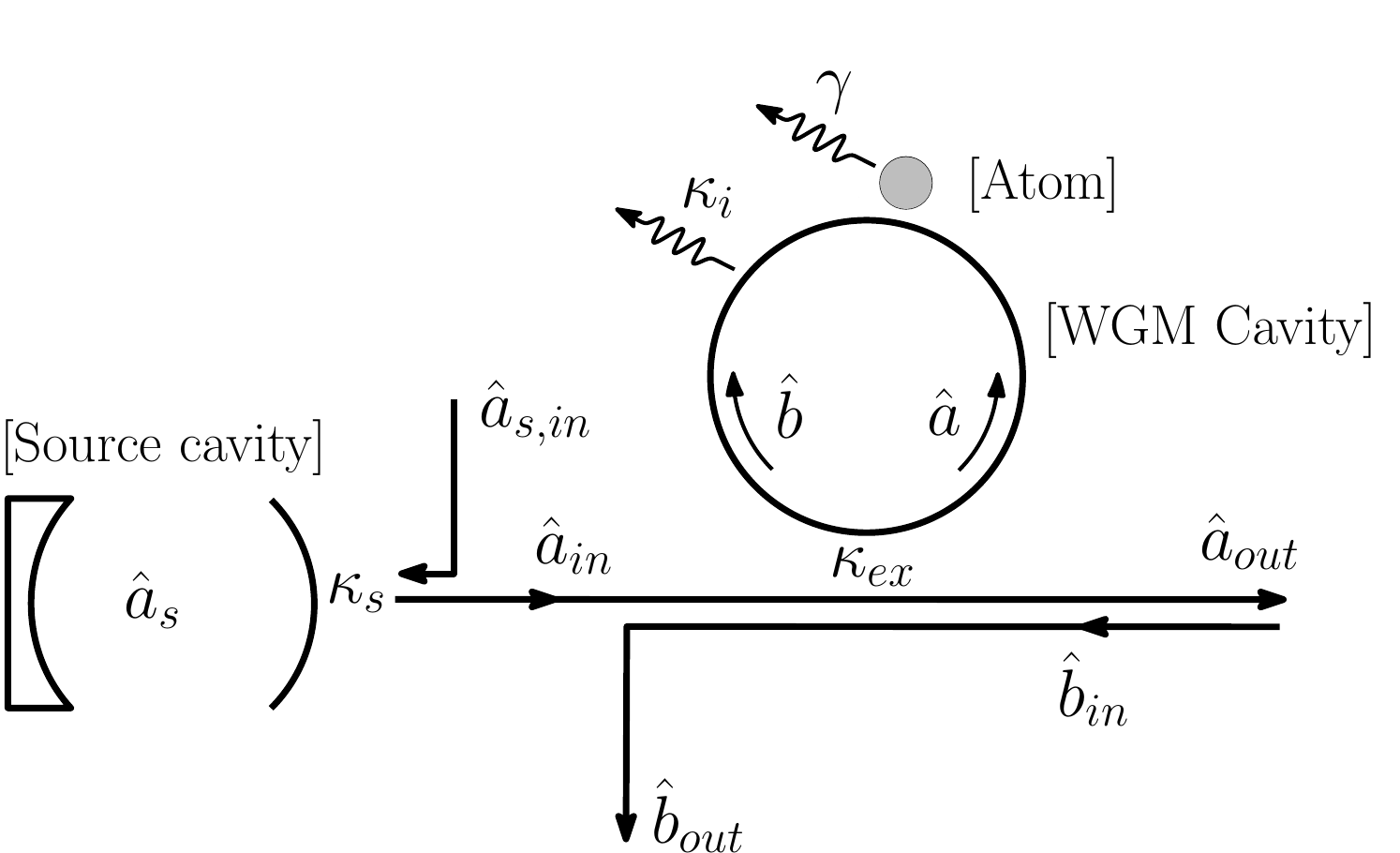}
\end{center}
\caption{Schematic of the theoretical model.}
\label{fig:scheme}
\end{figure}

The strong atom-photon interaction necessary for SPRINT can be achieved by means of whisperig gallery mode (WGM) microresonators~\cite{armani2003}, in which photons are confined by continuous total internal reflection for an extended period of time inside a very small volume, to the extent that the electric field of even a single photon is sufficient to significantly affect the dynamics of a nearby atom. The microresonator is nanofiber-coupled, so that photons coming from the left and right in the fiber excite the counterpropagating modes through evanescent-wave coupling (Fig.~\ref{fig:scheme}). In Ref.~\cite{shomroni2014} transverse magnetic (TM) modes were used, which have the remarkable property that their evanescent-wave polarization is circular to a high degree~\cite{junge2013,petersen2014}, with the handedness of the polarization depending on the direction of propagation. This creates the unique situation in which the counterclockwise ($a$) and clockwise ($b$) rotating photons may interact with different atomic transitions due to their opposite spin. In particular, in the case of a $\Lambda$-configuration atom in which the two ground-states and the excited state are associated with different total angular momentum, $a$ interacts only with the $\sigma^+$ transition between the left ground state $G_1$ and the excited state $e$, whereas $b$ interacts only with the $\sigma^-$ transition associated with the right ground state $G_2$ (Fig.\ref{fig:principles_new}b).
Since parasitic scattering between $a$ and $b$ is assumed to be negligible, the only way a photon can be reflected, namely to transfer from $a$ to $b$ or vice versa, is by Raman passage of the atom between its left and right ground states through the excited state. Note that another possible configuration for attaining the same physical situation consists of a a single-sided cavity supporting two orthogonal polarizations~\cite{rosenblum2011}.

\section{\label{sec:level2}Theoretical model and the influence of losses}

The mechanism of SPRINT is most conveniently modeled using quantum trajectory theory with cascaded systems~\cite{carmichael1993,rosenblum2011}. A single-sided source cavity (described by its annihilation operator $\hat{a}_s$) with a decay rate of $2\kappa_s$ emits a single photon to the right, after which it couples to the $a$ mode of the WGM microresonator at a rate $2\kappa_{ex}$. The photon then interacts with the atom's left and right ground states (through the excited state) at rates $g_1$ and $g_2$ respectively, and eventually leaks back into the nanofiber as a right or left propagating photon, described by the output field annihilation operators~\cite{gardiner1985}

\begin{subequations}
\label{output}
\begin{eqnarray}
\hat{a}_{out} &=& \sqrt{2\kappa_s}\hat{a}_s+ \sqrt{2\kappa_{ex}}\hat{a}\label{aout}\\
\hat{b}_{out} &=& \sqrt{2\kappa_{ex}}\hat{b}\label{bout},
\end{eqnarray}
\end{subequations}
where the vacuum input operators $\hat{a}_{s,in}$ and $\hat{b}_{in}$ have been discarded, since we only consider normally ordered operators. Equation~\ref{aout} exhibits the fact that a transmitted output photon cannot be exclusively attributed to the event of a photon emitted by the microresonator to the right, or to a transmitted photon that did not enter the microresonator in the first place. Rather, it is the interference between both that creates the resulting output field.
The Hamiltonian that governs the dynamics of this system contains a non-unitary term that describes the unidirectional driving of the resonator by the source cavity~\cite{carmichael1993}, loss and detuning terms, and two Jaynes-Cummings terms that describe the atom-field interaction (setting $\hbar=1$):
\begin{align}
\label{hamiltonian0}
\hat{H}_0=& - 2i\sqrt{\kappa_s\kappa_{ex}}\hat{a}_s\hat{a}^\dag-i\kappa_s \hat{a}_s^\dag \hat{a}_s  \nonumber\\
 &-i(\kappa+i\delta_C) ( \hat{a}^\dag \hat{a}+\hat{b}^\dag \hat{b})-i(\gamma+i\delta_a)\hat{\sigma}_{ee}\nonumber\\
&+(g_1^\ast\hat{a}^\dag\hat{\sigma}_{1e}+g_1\hat{\sigma}_{1e}^\dag\hat{a})+(g_2\hat{b}^\dag\hat{\sigma}_{2e}+g_2^\ast\hat{\sigma}_{2e}^\dag\hat{b}),
\end{align}
where $\hat{\sigma}_{ke}$ is the lowering operator from the excited state to ground state $G_k$, $\hat{\sigma}_{ee}$ is the excited state population, $2\kappa=2\kappa_{ex}+2\kappa_i$ is the total WGM loss rate due to nanofiber coupling and intrinsic loss, respectively, and $2\gamma$ is the atomic spontaneous emission rate into free space. The detuning of the atomic transitions  $\delta_a$, and the detuning of the cavity modes $\delta_C$ with respect to the driving frequency are taken to be zero unless otherwise specified. The two counterpropagating WGMs are degenerate by symmetry, and the two ground states are assumed to be degenerate as well. However, this degeneracy is not required for SPRINT as long as every mode is resonant with one of the transitions.

The initial state $\left|\psi(0)\right\rangle=\left|1_s 0_a 0_b G_1\right\rangle$ containing a photon in the source cavity and an atom in the left ground state evolves according to the Schr\"{o}dinger equation to
\begin{align}
\left|\psi(t)\right\rangle=&e^{-\kappa_s t}\left|1_s 0_a 0_b G_1\right\rangle + \alpha(t)\left|0_s 1_a 0_b G_1\right\rangle\nonumber\\
&+\beta(t)\left|0_s 0_a 1_b G_2\right\rangle+\xi(t)\left|0_s 0_a 0_b e\right\rangle
\end{align}
with
\begin{align}
\label{hamiltonian1}
\dot{\alpha}=& - 2\sqrt{\kappa_s\kappa_{ex}}e^{-\kappa_s t}-ig_1^\ast\xi-\kappa \alpha\nonumber\\
\dot{\beta}=& -ig_2\xi-\kappa \beta\nonumber\\
\dot{\xi}=& -ig_1 \alpha -ig_2^\ast \beta - \gamma\xi.
\end{align}
Ideally one uses very long pulses, so that $\kappa_s$ becomes the lowest rate of the system. $\left|\psi(t)\right\rangle$ is then close to steady-state at all times, and the derivatives can be put equal to zero, yielding

\begin{align}
\alpha=& - 2 \frac{\sqrt{\kappa_s\kappa_{ex}}}{\kappa}\left[1-\frac{|g_1|^2}{|g_1|^2+|g_2|^2}\frac{2C_{tot}}{1+2C_{tot}}\right] e^{-\kappa_s t}\nonumber\\
\beta=&2 \frac{\sqrt{\kappa_s\kappa_{ex}}}{\kappa}\frac{g_1 g_2}{|g_1|^2+|g_2|^2}\frac{2C_{tot}}{1+2C_{tot}}e^{-\kappa_s t}\nonumber\\
\xi=& 2i\sqrt{\kappa_s\kappa_{ex}}\frac{g_1}{|g_1|^2+|g_2|^2}\frac{2C_{tot}}{1+2C_{tot}}e^{-\kappa_s t},
\end{align}
where the total cooperativity $C_{tot}=(|g_1|^2+|g_2|^2)/2\kappa\gamma$ quantifies the tendency of the atom to emit into both microresonator modes, rather than into free space.
In this long-pulse limit, the fact that the source-cavity formalism results in an exponentially decaying input pulse is irrelevant, and the results below apply for any pulse shape. Using Eq.~\ref{output} the transmission and reflection probabilities can be calculated, giving
\begin{subequations}
\label{efficiency}
\begin{align}
T =& \int\limits_0^\infty \langle \hat{a}_{out}^\dag\hat{a}_{out} \rangle\, dt =\left|\frac{\kappa_{ex}}{\kappa}\frac{2|g_1|^2}{|g_1|^2+|g_2|^2}\frac{2C_{tot}}{1+2C_{tot}}+t_0 \right|^2\\
R =& \int\limits_0^\infty \langle \hat{b}_{out}^\dag\hat{b}_{out} \rangle\, dt
=\left|\frac{\kappa_{ex}}{\kappa}\frac{2g_1 g_2}{|g_1|^2+|g_2|^2}\frac{2C_{tot}}{1+2C_{tot}}\right|^2,
\end{align}
\end{subequations}
where $t_0=-\frac{\kappa_{ex}-\kappa_i}{\kappa}$ is the forward transmission when the atom is absent.
Inspection of Eqs.~\ref{efficiency} reveals three requirements for efficient operation of SPRINT, i.e. for $R$ close to unity: First, the coupling strengths of the two transitions must be equal in their absolute values~\cite{Zeeman_note}. Secondly, the intrinsic loss of the microresonator must be considerably smaller than $\kappa_{ex}$. Finally, the cooperativity must be significantly larger than one, ensuring that the spontaneous emission is primarily directed into the mircroresonator, rather than into free space. This situation can be realized both in the strong coupling regime (in which $g\gg\kappa_{ex}\gg\kappa_i,\gamma$) and in the fast-cavity or Purcell regime (with $\kappa_{ex}\gg g \gg\kappa_i,\gamma$).

In realistic systems photon losses may be present either due to intrinsic resonator loss, or because of limited cooperativity. But remarkably, even in this case the SPRINT fidelity, defined as $R/(R+T)$, can still be unity by choosing
\begin{align}
\label{loss}
\kappa_{ex}= \kappa_i\sqrt{1+2C_i},
\end{align}
where we took $g_1=g_2\equiv g$, and defined the total intrinsic cooperativity $C_i=|g|^2/\kappa_i\gamma$. Hence, by tuning the coupling strength (for example, by varying the distance between the microresonator and the nanofiber), one can ensure that complete destructive interference is maintained in the transmission. The photon can then either be reflected, or lost due to dissipation, but never transmitted - allowing heralded operation of SPRINT. One can gain more insight by noting that for large enough intrinsic cooperativity Eq.~\ref{loss} becomes $2\kappa_i/\kappa^2=\gamma/|g|^2$, suggesting that the loss rate in the cavity should compensate for the loss rate in the atom. Alternatively, from Eq.~\ref{efficiency}a, the forward radiation of the atom and the forward transmission when the atom is absent should have equal magnitude, but opposite sign.

\section{\label{sec:level3}Multiple excited states}

In practical systems, the presence of various excited states (See Fig.~\ref{fig:multilevel}a) can create multiple pathways for Raman passage, which interfere with each other, and therefore affect the operation of SPRINT.
This situation can be analyzed by adding to the Hamiltonian of Eq.~\ref{hamiltonian0} two Jaynes-Cummings terms corresponding to the second excited state $e'$, along with its associated loss $\gamma'$ and detuning $\delta_a'$:
\begin{align}
\label{hamiltonian1}
\hat{H}_1=&\hat{H}_0-i(\gamma'+i\delta_a')\hat{\sigma}_{e'e'}\nonumber\\
&+(g_1^{\prime\ast}\hat{a}^\dag\hat{\sigma}_{1e'}+g'_1\hat{\sigma}_{1e'}^\dag\hat{a})+(g'_2\hat{b}^\dag\hat{\sigma}_{2e'}+g_2^{\prime\ast}\hat{\sigma}_{2e'}^\dag\hat{b}).
\end{align}
The Schr\"{o}dinger equations corresponding to this new Hamiltonian can be easily solved by assuming $g_1'=\eta g_1$ and $g_2'= s\eta g_2$, with $s=\pm1$.
In the case of coupling strengths of equal sign, i.e. $s=+1$, we obtain
\begin{subequations}
\begin{align}
T =& \left|\frac{\kappa_{ex}}{\kappa}\frac{2|g_1|^2}{|g_1|^2+|g_2|^2}\frac{2(C_{tot}+C'_{tot})}{1+2(C_{tot}+C'_{tot})}+t_0  \right|^2\\
R =& \left|\frac{\kappa_{ex}}{\kappa}\frac{2g_1 g_2}{|g_1|^2+|g_2|^2}\frac{2(C_{tot}+C'_{tot})}{1+2(C_{tot}+C'_{tot})}\right|^2,
\end{align}
\end{subequations}
where $C'_{tot}=(|g_1'|^2+|g_2'|^2)/2\kappa(\gamma'+i\delta_a')$ is the complex cooperativity associated with the detuned second excited state. Not surprisingly, since the symmetry between the coupling strengths of both sides is maintained, the four-level system effectively behaves like a symmetric $\Lambda$-system with cooperativity equal to the sum of the cooperativities of the separate transitions.
However, if $s=-1$ we obtain
\begin{subequations}
\begin{align}
T =& \left|\frac{\kappa_{ex}}{\kappa}\frac{2|g_1|^2}{|g_1|^2+|g_2|^2}\frac{2(C_{tot}+C'_{tot})+16\frac{|g_1|^2+|g_2|^2}{|g_1|^2}C_1C_2'}{1+2(C_{tot}+C'_{tot})+16C_1C_2'}\right.
\nonumber\\
&\left.+t_0  \right|^2\label{constructive}\\
R =& \left|\frac{\kappa_{ex}}{\kappa}\frac{2g_1 g_2}{|g_1|^2+|g_2|^2}\frac{2(C_{tot}-C'_{tot})}{1+2(C_{tot}+C'_{tot})+16C_1C_2'}\right|^2\label{destructive},
\end{align}
\end{subequations}
where we defined single-transition cooperativities $C_1=|g_1|^2/2\kappa\gamma$ and $C_2'=|g_2'|^2/2\kappa(\gamma'+i\delta_a')$. In this case, the symmetry between the two `sides' of the system is broken, removing the balance necessary for SPRINT. Still, unit fidelity can be attained by slightly detuning the cavity. The necessary detuning and fiber-resonator coupling can be determined by replacing $\kappa_i\rightarrow \kappa_i+i\delta_C$ in Eq.~\ref{constructive}, and setting $T=0$ for both the real and imaginary parts.
A useful approximation is then obtained by taking $\gamma'\ll i \delta_a'$, reflecting the fact that the detuning of the second excited state is high compared to its linewidth. The optimal parameters then become
\begin{subequations}
\begin{align}
\delta_C=&\kappa_i C_i'\frac{1+2C_i}{1+C_i}\\
\kappa_{ex}=& \kappa_i\sqrt{\left(1+2C_i'\frac{\gamma'}{\delta_a'}+\frac{C_i'^2}{(1+C_i)^2}\right)(1+2C_i)},
\end{align}
\end{subequations}
where we took $g_1=g_2\equiv g$, and $g_1'=-g_2'\equiv g'$ and defined the total intrinsic cooperativity of the detuned excited state $C_i'=|g'|^2/\kappa_i\delta_a'$.

\section{\label{sec:level4}Simulations}

\begin{figure}
\begin{center}
\includegraphics[width=0.45\textwidth]{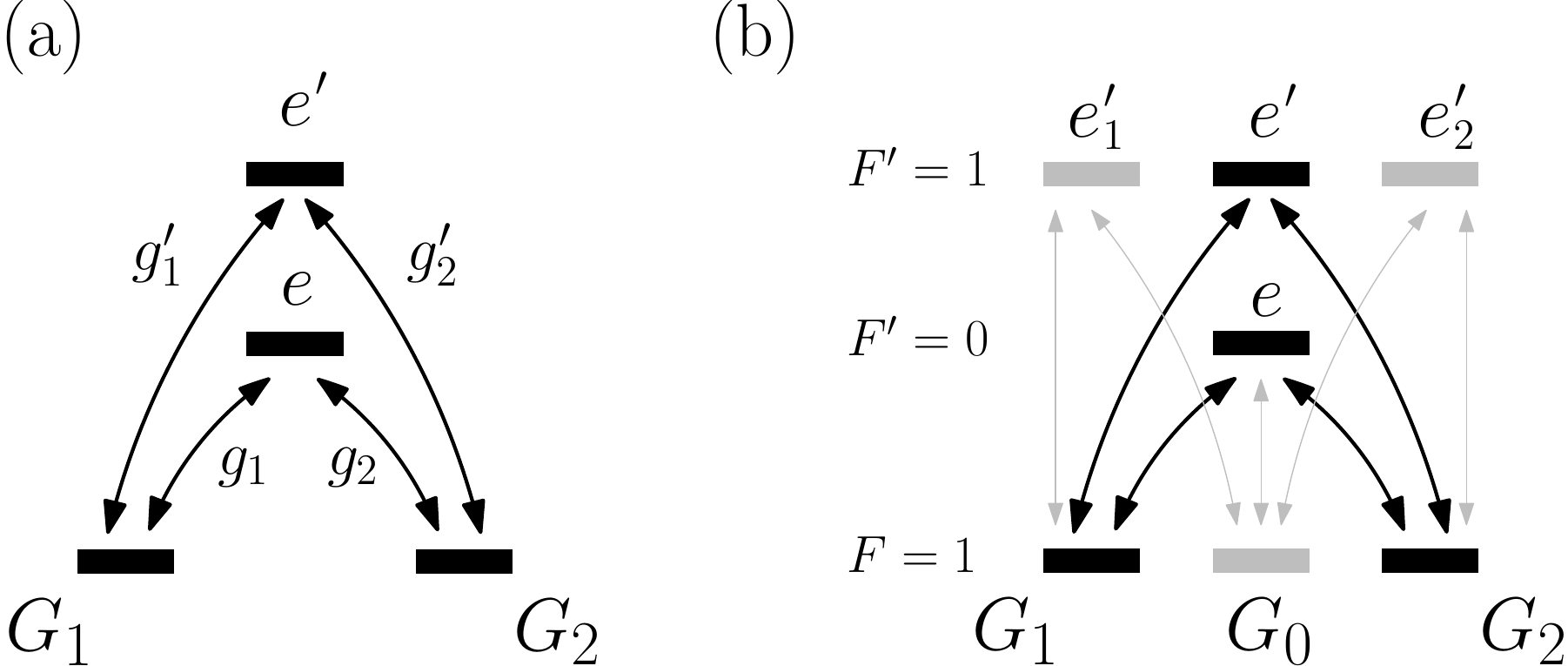}
\end{center}
\caption{ (a) Two excited states instead of one provide two pathways for SPRINT. Depending on the signs of the coupling strengths, the symmetry required for SPRINT can be either maintained or removed. (b) Actual level scheme of the $^{87}$Rb $D_2$ line. Transitions that are weak due to polarization mismatch are shown in grey. Note that the transition $G_0\rightarrow e'$ is forbidden, and that the effect of the excited state $F'=2$ is negligible, due to its large detuning and low coupling strength to $F=1$.}
\label{fig:multilevel}
\end{figure}

In this section we analyze the imperfections of SPRINT specific to its realization in Ref.~\cite{shomroni2014}, which uses $^{87}$Rb atoms coupled to silica microsphere resonators. A major hurdle is the fact that trapping atoms near WGM microresonators is a highly challenging task that has not yet been realized. As a result, the atoms fly past the microresonator or crash into its surface, causing the coupling strength to vary from run to run, and during a single run. This can be taken into account by assuming a normally distributed coupling strength with independently determined mean $\bar{g}$ and standard deviation $\sigma_g$, truncated at the minimally detectable and maximally available coupling strengths. Another effect is the nonideal circular polarization in the evanescent part of the WGM. The analytical solution of Maxwell's equations for a microsphere~\cite{johnson1993} show the existence of a non-negligible $\pi$-polarized (i.e. perpendicular to the WGM plane) electric field, and also a component of circular polarization with opposite handedness. At the location where the atom feels the strongest field, calculation of the ratios of the unwanted electric field components to the desired ones for a $20~\mu$m-radius microsphere yields $r_\sigma\simeq 0.18$ for the opposite circular polarized field, and  $r_\pi\simeq 0.13$ for the $\pi$-polarized field. This impurity may for example cause a photon in the $b$ mode to drive a $\sigma^+$ transition of the atom, disrupting the SPRINT mechanism. The presence of these fields also entails that otherwise uncoupled atomic levels can now take part in the dynamics. In $^{87}$Rb, all three ground states of $F=1$ should now be taken into account (See Fig.~\ref{fig:multilevel}b), as well as all three excited states in $F'=1$.
Rayleigh scattering between modes $a$ and $b$ at a rate $2h$ can also affect the fidelity in various ways. First, a photon can be reflected without the atom being involved. Secondly, nonzero $h$ causes the formation of an azimuthally-varying field. Since every atomic transition interacts with both $a$ and $b$ due to the nonideal circular polarization, the azimuthal location of the atom, represented by the phase of the coupling constants, starts to play a role. Finally, the effect of different pulse lengths and shapes can be included by introducing a time-varying $\kappa_s$.
All these effects can be taken into account by simulating the evolution of the state of the system with the complete Hamiltonian, including all transitions of interest in $^{87}$Rb:

\begin{align}
\label{FullH}
\hat{H}_2 =& \hat{H}_1-i(\gamma'+i\delta_a')\hat{\sigma}_{e_1'e_1'}-i(\gamma'+i\delta_a')\hat{\sigma}_{e_2'e_2'}+h(\hat{a}^\dag\hat{b}+\hat{b}^\dag\hat{a})\nonumber\\
&+r_\sigma(g^{\ast}\hat{a}^\dag\hat{\sigma}_{2e}+g\hat{\sigma}_{2e}^\dag\hat{a})
+r_\sigma(g\hat{b}^\dag\hat{\sigma}_{1e}+g^{\ast}\hat{\sigma}_{1e}^\dag\hat{b})\nonumber\\
&-r_\sigma(g^{\prime\ast}\hat{a}^\dag\hat{\sigma}_{2e'}+g'\hat{\sigma}_{2e'}^\dag\hat{a})
+r_\sigma(g'\hat{b}^\dag\hat{\sigma}_{1e'}+g^{\prime\ast}\hat{\sigma}_{1e'}^\dag\hat{b})\nonumber\\
&+r_\pi(g^{\ast}\hat{a}^\dag\hat{\sigma}_{0e}+g\hat{\sigma}_{0e}^\dag\hat{a})
+r_\pi(g\hat{b}^\dag\hat{\sigma}_{0e}+g^{\ast}\hat{\sigma}_{0e}^\dag\hat{b})\nonumber\\
&-r_\pi(g^{\prime\ast}\hat{a}^\dag\hat{\sigma}_{1e_1'}+g'\hat{\sigma}_{1e_1'}^\dag\hat{a})
-r_\pi(g'\hat{b}^\dag\hat{\sigma}_{1e'_1}+g^{\prime\ast}\hat{\sigma}_{1e'_1}^\dag\hat{b})\nonumber\\
&+r_\pi(g^{\prime\ast}\hat{a}^\dag\hat{\sigma}_{2e_2'}+g'\hat{\sigma}_{2e_2'}^\dag\hat{a})
+r_\pi(g'\hat{b}^\dag\hat{\sigma}_{2e'_2}+g^{\prime\ast}\hat{\sigma}_{2e'_2}^\dag\hat{b})\nonumber\\
&+(g^{\prime\ast}\hat{a}^\dag\hat{\sigma}_{0e_2'}+g'\hat{\sigma}_{0e_2'}^\dag\hat{a})
-(g'\hat{b}^\dag\hat{\sigma}_{0e_1'}+g^{\prime\ast}\hat{\sigma}_{0e_1'}^\dag\hat{b})\nonumber\\
&-r_\sigma(g^{\prime\ast}\hat{a}^\dag\hat{\sigma}_{0e_1'}+g'\hat{\sigma}_{0e_1'}^\dag\hat{a})
+r_\sigma(g'\hat{b}^\dag\hat{\sigma}_{0e_2'}+g^{\prime\ast}\hat{\sigma}_{0e_2'}^\dag\hat{b}),
\end{align}

where we used the fact that in our case all coupling strengths to a given excited state manifold are of equal magnitude. The signs, however, vary, and in particular, one should set $g_1'= g'$ and $g_2'= -g'$ in $\hat{H}_1$, and $g_1= g_2=g$ in $\hat{H}_0$, resulting in the deterioration of SPRINT fidelity.
The best strategy is to work with the $F=1\rightarrow F'=1$ manifold on resonance, rather than with $F=1\rightarrow F'=0$. This serves a double purpose: the efficiency of SPRINT can be increased, because $g'=\sqrt{5/4} g$, resulting in an increased cooperativity. For the same reason, the interference with the red-detuned $F=1\rightarrow F'=0$ transitions is reduced.

In order to assess the performance of SPRINT in an actual experimental setting, we simulated the dynamics using realistic system parameters $(\kappa_i,\bar{g},\sigma_g,h,\gamma,\gamma',\delta_a,\delta_a')=2\pi\times(6,16,6,1,3,3,0,-72)$ MHz \cite{shomroni2014}. Moreover, we used the optimal nanofiber-microresonator coupling rate and microresonator detuning $(\kappa_{ex},\delta_C)=2\pi\times(30,-7)$ MHz, a $53$ ns full width at half maximum Gaussian input pulse, and a uniformly distributed atomic azimuthal location. As shown in Table \ref{tab:G}a, a simulation using the Hamiltonian of Eq.~\ref{FullH} with the atom initialized in ground state $G_1$ results in an optimal SPRINT fidelity of $\sim88\%$, and a photon loss probability of $\sim 51\%$. A reflection event heralds a successful transfer of the atom to the opposite ground state with a probability of $\sim94\%$.
After a first photon toggled the atom to $G_2$, a second photon that is sent from the source cavity still has some probability of being reflected, due to the nonideal circular polarization. A simulation with the atom initially in $G_2$ yields a reflection probability of $\sim3.5\%$, but this reflection no longer induces the Raman passage of the atom.

\begin{center}
    \begin{tabular}{ | m{2cm}  | >{\centering\arraybackslash}m{1.5cm}  | >{\centering\arraybackslash}m{1.5cm}  | >{\centering\arraybackslash}m{1.5cm}  | >{\centering\arraybackslash}m{1.5cm}  |}
    \multicolumn{1}{c|}{(a)}    & R    & T    & L   &\multicolumn{1}{c|}{Total} \\ \hline
    \multicolumn{1}{c|}{Toggle}    & 39.97\,\% & 1.39\,\% & 17.03\,\% &58.39\,\% \\ \hline
    \multicolumn{1}{c|}{No toggle}  & 2.17\,\%  & 4.18\,\% & 30.15\,\% &36.50\,\% \\ \hline
    \multicolumn{1}{c|}{Atom lost}  & 0.50\,\%  & 0.43\,\% & 4.16\,\%  & 5.09\,\% \\ \hline
    \multicolumn{1}{c|}{Total} & 42.64\,\% & 6.00\,\% & 51.34\,\%    & 100\,\%\\ \hline
    \end{tabular}
\end{center}
\vspace{.5cm}
\begin{center}
    \begin{tabular}{ | m{2cm}  | >{\centering\arraybackslash}m{1.5cm}  | >{\centering\arraybackslash}m{1.5cm}  | >{\centering\arraybackslash}m{1.5cm}  | >{\centering\arraybackslash}m{1.5cm}  |}

    \multicolumn{1}{c|}{(b)}    & R    & T    & L   &\multicolumn{1}{c|}{Total} \\ \hline
    \multicolumn{1}{c|}{Toggle} & 0.06\,\% & 1.39\,\% & 0.59\,\%&2.04\,\% \\ \hline
    \multicolumn{1}{c|}{No toggle}& 1.55\,\% & 41.97\,\% & 51.72\,\%&95.24\,\% \\ \hline
    \multicolumn{1}{c|}{Atom lost} & 0.01\,\% & 1.76\,\% & 0.95\,\% & 2.75\,\% \\ \hline
    \multicolumn{1}{c|}{Total} & 1.62\,\% & 45.12\,\% & 53.26\,\%    & 100\,\%\\ \hline
    \end{tabular}
\end{center}

\captionof{table}{ \label{tab:G}Statistics of atomic and photonic final states for: (a) the atom initially in $G_1$, and (b), the atom initially in $G_2$. $R$ stands for probability of reflection, $T$ for probability of transmission, and $L$ for probability of photon loss. The atom can either toggle to the other ground state, remain in the initial one, or end up in $G_0$ or in $F=2$ (not shown in Fig.~\ref{fig:multilevel}b), in which case it is considered lost.}

\section{\label{sec:level5}Conclusion}

In this work we analyzed the performance of single-photon Raman interaction, and showed its robustness against various experimental flaws. While atomic emission to free space, and intrinsic cavity losses unavoidably result in photon loss, an ideal fidelity can still be maintained by choosing the coupling rate between the nanofiber and the microresonator appropriately, albeit at the price of reduced efficiency. The effect of a second detuned routing pathway that disrupts the operation of SPRINT can be annulled by slightly detuning the cavity resonance. The analysis presented in this work provides the tools for realistic optimization of SPRINT, bringing it into the regime in which it could form the basis for a passive, all-optical quantum node.
\\

\acknowledgments
This work was partially supported by the Israel Science Foundation, the Wolfson Foundation and the Crown Photonics Center. This research was made possible in part by the historic generosity of the Harold Perlman Family.

\bibliography{mybib}{}

\end{document}